\documentclass[aps,prb,notitlepage,citeautoscript,superscriptaddress,eqsecnum,reprint, longbibliography]{revtex4-2}
\usepackage[english]{babel}

\usepackage{placeins}  

\usepackage{xr}
\usepackage{amsmath}
\usepackage{amsfonts, amssymb,amsxtra}
\usepackage[]{graphicx}
\usepackage{grffile}
\usepackage{amsfonts}
\usepackage{framed}
\usepackage{bm}
\usepackage{braket}
\usepackage{xcolor}
\usepackage{chngcntr}
\usepackage{comment}
\usepackage[normalem]{ulem}
\usepackage[linesnumbered,ruled,vlined]{algorithm2e} 
\usepackage[noend]{algpseudocode}
\counterwithout{equation}{section} 

\usepackage{bm}
\usepackage{epsfig}
\usepackage{blindtext}
\usepackage[most]{tcolorbox}
\usepackage{verbatim}
\usepackage{color}
\usepackage{diagbox}
\usepackage{outlines}
\usepackage{cancel}
\usepackage{simpler-wick}

\providecommand{\abs}[1]{\lvert#1\rvert}

\DeclareMathOperator*{\argmax}{argmax}

\usepackage[colorlinks=true]{hyperref}
\hypersetup{
    unicode=false,          
    pdftoolbar=true,        
    pdfmenubar=true,        
    pdffitwindow=false,     
    pdfstartview={FitH},    
    pdftitle={},    
    pdfauthor={},     
    pdfsubject={},   
    pdfcreator={},   
    pdfproducer={}, 
    pdfkeywords={} {} {}, 
    pdfnewwindow=true,      
    colorlinks=true,       
    linkcolor=magenta, 
    citecolor=blue,        
    filecolor=magenta,      
    urlcolor=blue           
}

\begin{document}
\title{Iterative Optimization with Partial Convergence Guarantees on Neutral Atom Quantum Computers}
\author{Cédrick Perron}\email{cedrick.perron@usherbrooke.ca}
\affiliation{Institut quantique, Sherbrooke, Québec, J1K 2R1, Canada}
\affiliation{Département de génie électrique et de génie informatique, Université de Sherbrooke, Sherbrooke, Québec, J1K 2R1, Canada}
\author{Yves Bérubé-Lauzière}\email{yves.berube-lauziere@usherbrooke.ca}
\affiliation{Institut quantique, Sherbrooke, Québec, J1K 2R1, Canada}
\affiliation{Département de génie électrique et de génie informatique, Université de Sherbrooke, Sherbrooke, Québec, J1K 2R1, Canada}
\author{Victor Drouin-Touchette}
\email{victor.drouin-touchette@usherbrooke.ca}
\affiliation{Institut quantique, Sherbrooke, Québec, J1K 2R1, Canada}
\affiliation{Département de génie électrique et de génie informatique, Université de Sherbrooke, Sherbrooke, Québec, J1K 2R1, Canada}
\date{\today}

\begin{abstract}
Neutral atom quantum computers (NAQCs) have emerged as a promising platform for solving the maximum weighted independent set (MWIS) problem. However, analog quantum approaches face two key limitations: Constraints of the atomic layout on realizable graph geometries and the absence of performance guarantees. We introduce \textsc{Lp-Quts}, a hybrid quantum-classical framework that integrates an NAQC sampler into a classical cutting-plane algorithm. At each iteration, a relaxed linear program (RLP) bounds the MWIS and induces a reduced graph from which independent sets are sampled using an analog quantum sampler. A novel sample-informed separation problem guides odd-cycle cut selection and accelerates convergence. For t-perfect graphs, \textsc{Lp-Quts} inherits polynomial-time convergence guarantees from the classical theory of cutting planes. We evaluate our approach on instances with up to 300 vertices—a scale that exceeds the capabilities of current NAQCs hardware. In this regime, \textsc{Lp-Quts} reaches solutions within 5–10\% of optimality, outperforming direct analog quantum protocols and greedy baselines under equal sampling budgets. As expected, simulated annealing remains the strongest sample-based solver at this scale. These results demonstrate how quantum samplers can be effectively embedded within classical optimization frameworks to deliver near-optimal solutions with reduced quantum resources while preserving formal guarantees.

\end{abstract}

\maketitle

%

\section{Introduction}
\label{sec:introduction}

Neutral atom quantum computers are a recent hardware modality~\cite{saffman_quantum_2010} with which both digital and analog quantum routines can be enacted \cite{henriet_quantum_2020}. Their development by both academic groups and private companies has led to demonstrations of error correction~\cite{bluvstein2026fault, sales2025experimental}, advanced quantum simulation~\cite{bluvstein_controlling_2021, semeghini2021probing, browaeys_many-body_2020, scholl_quantum_2021, ebadi_quantum_2021, gonzalez-cuadra_fermionic_2023}, machine learning~\cite{kornjaca_trimer_2022, henry_quantum_2021, leclerc2023financial} and optimization~\cite{ebadi_quantum_2022, dalyac2024graph} capabilities. At its simplest, a neutral atom quantum computer (NAQC) uses alkaline atoms -- typically Rubidium (Rb) -- that can be individually trapped at arbitrary positions $\{ \Vec{r}_i \}$ in a two-dimensional plane using optical tweezers and cooled to their electronic ground state ($|5S_{1/2}\rangle$ for Rb). In the analog mode of operation, quantum information is encoded in the two level system consisting of the ground state $|0\rangle = |5S_{1/2}\rangle$ and a highly excited Rydberg state $|1\rangle = |60S_{1/2}\rangle$ of each atom. Transitions between these states are driven by a pair of lasers, giving rise to a time-dependent field acting on the qubits \cite{henriet_quantum_2020}.

These atoms interact through repulsive van der Waals forces $V_{ij} \sim 1/r_{ij}^6$, where $r_{ij} = \lvert \Vec{r}_i - \Vec{r}_j \rvert$ is the distance between atoms. In the regime where the interaction strength exceeds the drive amplitude, $V_{ij} > \Omega$, simultaneous excitation of nearby atoms $i, j$ is prohibited, giving rise to the Rydberg blockade phenomenon. This can be used to embed graph structures on the NAQC \cite{dalyac2024graph}, where atoms represent vertices and edges are realized through the Rydberg blockade if atoms are closer than $R_b(\Omega)$, the blockade radius. Furthermore, the driving laser's detuning, $\delta_i$, can be tuned to implement weights on these vertices. 

Together, these two elements lead to a natural interpretation for the ground state of such a blockaded ensemble of atoms under a spatially modulated detuning: It represents the maximum weighted independent set (MWIS) problem on a weighted undirected graph $G = (V, E, w)$, where atoms are vertices $v \in V$, and the Rydberg blockade between two atoms $v,v'$ represents an edge $(v,v') \in E$. A MWIS problem consists in finding the subset of vertices $S \subseteq V$ such that no two vertices in $S$ are adjacent, i.e. $(v, v') \notin E$ for $v, v' \in S$, and such that the sum of their weights $\sum_{v \in S} w_v$ is maximized. This combinatorial problem is NP-complete \cite{Lucas_2014} and enjoys numerous industrial applications \cite{wurtz2022industry}, notably in logistics and telecommunications. 
Indeed, recent experiments indicate that NAQCs can tackle MWIS instances on graphs with ten to hundreds of nodes \cite{hard-mwis-instances, wurtz2023aquilaqueras256qubitneutralatom}, with a pathway towards thousands of atoms \cite{dalyac2024graph, manetsch2025tweezer}. With their long coherence time and reconfigurable connectivity between atoms, they offer a promising route for addressing problems that are challenging for current classical heuristics \cite{hard-mwis-instances, ebadi_quantum_2022}.

However, realizing this potential remains fraught with challenges. Most real-world problems exhibit non-trivial connectivity that cannot be directly embedded into the native geometry of graphs accessible to NAQCs, namely, unit-disk (UD) graphs. One workaround is the use of gadgets, which encode general graphs into UD form by introducing chains of ancilla atoms \cite{ babbush_resource_2013, kim_rydberg_2022, byun_finding_2022, dalyac_exploring_2023,
nguyen_quantum_2023, lanthaler_rydberg-blockade-based_2023, jeong_quantum_2023, bombieri2025quantum}. These gadget-based reductions preserve a one-to-one correspondence between the optimal solution (i.e. the MWIS) of the original graph and the UD-reduced one. However, this comes with significant costs. First, determining the optimal number of ancillas for the UD-reduced graph amounts to solving a set cover problem, which is NP-Complete \cite{babbush_resource_2013}. Although polynomial-time heuristics to approximate this step exists, the embeddings in such cases are guaranteed to incur at least a quadratic overhead in the number of atoms $\mathcal{O}(|V|^2)$, where $|V|$ is the number of vertices of the graph \cite{cazals_quantum_2025}. Second, these reductions do not preserve the spectral structure of near-optimal solutions: Solutions of the UD-reduced instance with small optimality gaps map back to solutions of the original graph with quality comparable to random sampling \cite{cazals_quantum_2025}. This limitation is particularly consequential given that, for large-scale instances, obtaining the exact ground state of the UD-reduced graph is typically infeasible, as the adiabatic quantum optimization algorithm \cite{farhi2000quantum} is still expected to require runtimes scaling exponentially with system size due to the spectral gap closing \cite{Hen2013Complexity}. This renders approximate solutions the only realistic objective in practical scenarios. Alternatively, heuristic embedding strategies offer a practical route to approximate general graphs into a UD geometry \cite{cazals_quantum_2025, coelho_efficient_2022,  perron2025leveraginganalogneutralatom, schuetz2025quantum}. However, the quality of the resulting solutions is tightly linked to the fidelity of the UD approximation, which typically deteriorates with increasing system size and connectivity \cite{perron2025leveraginganalogneutralatom}. While single bitflip greedy post-processing techniques can modestly improve performance, its effectiveness diminishes as problem size grows unless additional heuristic rules or decision layers are incorporated \cite{Taillard2023HeuristicAlgorithms}.

Adding to these challenges, most quantum heuristics for the MWIS, including those based on quantum annealing, lack provable performance bounds on general graphs. Some exceptions exist, such as QAOA on 3-regular graphs \cite{farhi2014qaoa,farhi2019quantum}. However, it remains generally unclear for which classes of graphs quantum heuristics can reliably outperform classical ones. Additionally, current quantum solvers rarely leverage the extensive toolkit assembled by operations research (OR) researchers. Over decades, this field, which is the mathematical science of optimal decision-making, developed essential techniques, such as linear programming~\cite{dantzig2002linear}, integer programming~\cite{wolsey2020integer}, branching~\cite{morrison2016branch}, cutting-planes~\cite{applegate1995finding, desaulniers_cutting_2011}, and column generation~\cite{desrosiers2024branch}, that are all designed to tackle large-scale real-world optimization problems, often by identifying efficient decompositions. Without exploiting these principles, quantum solvers are thus typically unable to provide the same performance guarantees on the instances that are easy for classical algorithms. Consequently, the practical use of quantum heuristics at industrial scale remains difficult to justify.

\begin{figure*}[t!]
  \centering
  \includegraphics[width = 0.8\linewidth]{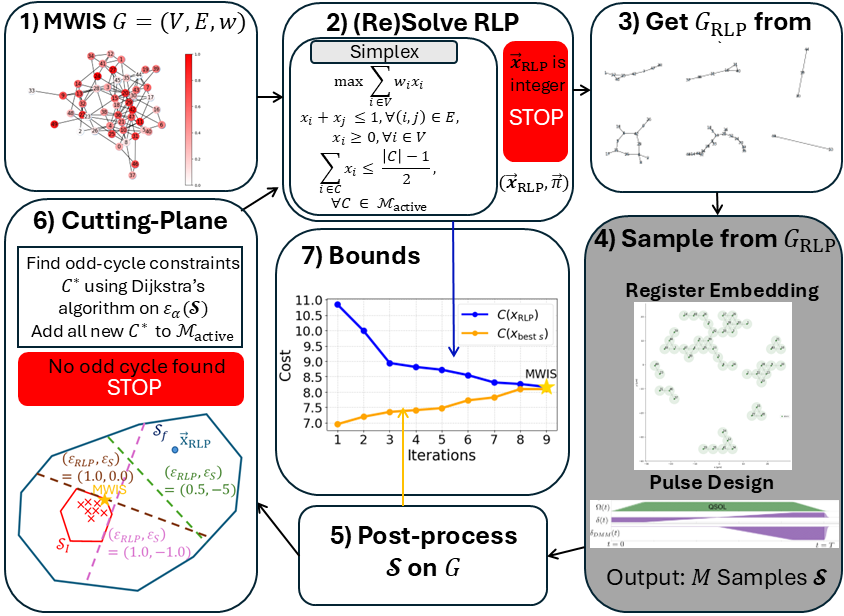}
  \caption{Overview of \textsc{Lp-Quts}. The algorithm iterates steps 2--6 to  tighten the upper (blue - RLP solutions) and lower (yellow - best independent set sampled) bounds on the MWIS, as shown in step 7). After the RLP is solved in step 2), we use the dual solution $\Vec{\pi}$ to generate the reduced graph $G_{\mathrm{RLP}}$ where edges correspond to all the tight inequalities in the optimal solution to the RLP. Then, in step 4), this graph is embedded on a NAQC and a laser pulse is used to sample independent sets. After postprocessing the samples $\mathcal{S}$ on the original graph $G$ in step 5), the bounds are improved step 7), and a separation problem is solved to identify a new cutting plane, i.e. an odd-cycle, for the RLP (step 6). The bottom panel of 6) illustrates our sample-informed separation problem. $\mathcal{S}_t$ represents the space of all possible solutions with $\vec{x}_{\mathrm{RLP}}$ as the current best one. Green, pink, and brown cuts separate this solution from the integer hull ($\mathcal{S}_I$; red), with sampled solutions marked by red ``X''s. The RLP cost $\varepsilon_{\mathrm{RLP}}$ distinguishes the worst (green) cut but cannot separate the pink and brown cuts. Incorporating the sampler-based cost $\varepsilon_s(\mathcal{S})$ favors the brown cut because it is in closer proximity to the sampled solutions. The brown cut is facet-defining and resolves the MWIS. 
  } 
  \label{fig:algo}
\end{figure*}

In this paper, we bridge this gap by introducing \textsc{Lp-Quts}, a hybrid quantum-classical iterative framework for solving MWIS problems on general graphs, integrating a quantum independent-set sampler into a classical cutting-plane algorithm. This algorithm is presented in Fig.~\ref{fig:algo}. We present the methodology in detail in Sec.~\ref{sec:methods}, and results in Sec.~\ref{sec:results-discussion}. The algorithm begins by solving the relaxed linear program (RLP), whose solution provides an upper bound on the MWIS and is used to construct a reduced graph that often decomposes into smaller and sparser subgraphs. Independent sets are sampled from this reduced graph using a quantum sampler and lifted to the original graph via greedy post-processing, providing a lower bound. Borrowing from cutting-plane algorithms, we then use these samples to guide the selection of odd-cycle inequalities added to the RLP, favoring cuts closer to the sampled solutions. Iterating this process tightens both bounds simultaneously: the RLP converges toward the MWIS (upper bound), while the improved reduced graph yields better samples (lower bound). Moreover, for t-perfect graphs \cite{chudnovsky_tperfect_graphs, bruhn2011clawfree}, our algorithm also inherits the same polynomial time (i.e. $\text{poly}(N)$ time, where $N$ is the number of vertices) convergence guarantees as classical odd-cycles cutting-plane \cite{groetschel1988_geometric, mahjoub2004_chap_polyedres}.

\begin{figure}[t!]
  \centering
  \includegraphics[width = 0.8\linewidth]{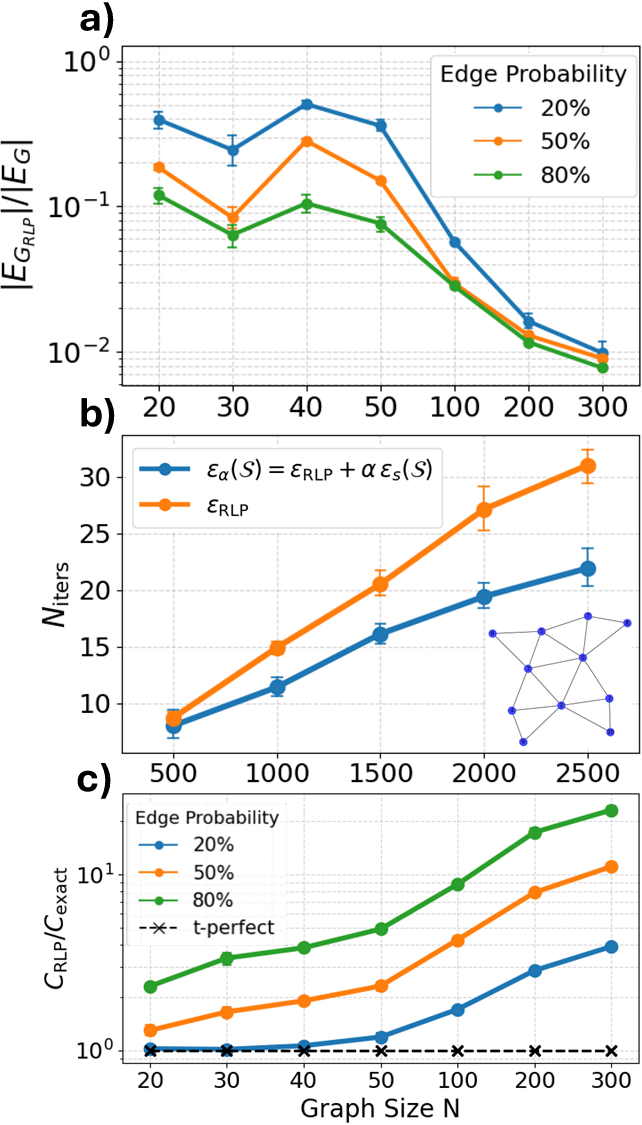}
  \caption{a) Ratio of the number of edges in the final $G_{\mathrm{RLP}}$ to that of the original graph $G$ for varying graph sizes $N$ and edge densities $p$. Each data point is the average of 10 random instances. b) Number of iterations to convergence for series--parallel graphs of different sizes (an example of such graph is shown in the inset). Series--parallel graphs are examples of t-perfect graphs and we observe polynomial scaling with system size, as expected, while seeing that our modified separation cost (blue) converges faster than the standard approach (orange), with the performance gap widening for larger graphs. c) Final optimality gap between the RLP solution and the true optimum. For t-perfect graphs, convergence to the optimal solution for all system sizes is observed as expected. For general graphs, the gap grows exponentially with system size and is larger for denser graphs. This reflects that odd-cycle inequalities alone are not sufficient in this general case, and additional cutting planes inequalities are required.
  } 
  \label{fig:perfideal}
\end{figure}

\section{Methods}
\label{sec:methods}

\textsc{Lp-Quts} borrows most of the classical aspects of its structure from the branch-and-cut algorithms of Refs.~\cite{barthel_clustering_2004, weigt2000number, hartmann2012phase,dewenter_phase_2012,claussen2022cutting}, where a similar classical algorithm was developed, characterized and used to study the hardness transition~\cite{moore2011nature} in the MWIS problem as a function of edge density. Similarly, the idea of combining decomposition tools and techniques from OR with a quantum sampler has resulted in a flurry of hybrid algorithms~\cite{brady2023iterative, maciejewski2024improving, dupont2023quantum, da_silva_coelho_quantum_2023, naghmouchi2024mixed, hirama2023efficient, maciejewski2024multilevel, czegel2025quantum, finvzgar2023quantum}. In this paper, we combine these elements, specifically targeting the optimization of MWIS problems with partial convergence guarantees using NAQCs. The following paragraphs detail the workflow presented in Fig.~\ref{fig:algo}.

\textit{Linear Programming}: \textsc{Lp-Quts} starts by formulating the MWIS problem on a general weighted graph $G = (V,E,w)$, where $V$ and $E$ are the numbers of vertices and edges of the graph respectively, as an integer linear program \cite{papadimitriou1998combinatorial}:
\begin{subequations} \label{eq-psp}
\begin{align}
    &\max \sum_{i\in V} w_i x_i, \label{eq-psp1}\\
    &x_i + x_j \le 1,\quad \forall (i,j)\in E, \label{eq-psp2}\\
    &x_i \in \{0,1\},\quad \forall i\in V, \label{eq-psp3}
\end{align} 
\end{subequations} 
where $x_i = 1$ if vertex $i$ with weight $w_i$ is included in the independent set, and $x_i = 0$ otherwise. Constraint \eqref{eq-psp2} ensures that adjacent vertices cannot both be selected. Due to the combinatorial hardness of the problem, the binary constraints \eqref{eq-psp3} are relaxed to $0 \leq x_i \leq 1$ to produce a relaxed linear program (RLP) that can be solved efficiently using the simplex method \cite{dantzig1963linear, papadimitriou1998combinatorial}, generally in $\text{poly}(V \cdot E)$ time. The relaxation tends to produce \textit{fractional/incomplete solutions} containing some fractional or undefined variables $x_i \in (0,1)$. The objective value $\sum_i w_i x_i$ of the relaxed solution provides an upper bound on the integer solution, and if all its variables are integer-valued, it is the MWIS. In \textsc{Lp-Quts}, the RLP is solved \footnote{We use the  GLPK solver \cite{glpk}} (see step 2 of Fig.~\ref{fig:algo}), which returns the solution $\Vec{x}_{RLP}$ as well as the solution $\Vec{\pi}$ of its corresponding \textit{dual problem}:
\begin{subequations}\label{eq:dual}
\begin{align}
    &\min \sum_{(i,j)\in E} \pi_{ij}, \label{eq:dual1} \\
    &\sum_{j:(i,j)\in E} \pi_{ij} \ge w_i, \quad \forall i \in V, \label{eq:dual2} \\
    &\pi_{ij} \ge 0, \quad \forall (i,j)\in E, \label{eq:dual3}
\end{align}
\end{subequations}
where each $\pi_{ij}$ represents the dual variable associated with each corresponding constraint \eqref{eq-psp2} of the RLP. At the optimal solution, one finds that each dual variable $\pi_{ij}$ is positive only when its corresponding constraint \eqref{eq-psp2} is tight $(x_i + x_j = 1)$, a condition known as \textit{complementary slackness} \cite{bertsimas1997introduction}. This reflects how strongly this constraint is important in shaping the structure of the current RLP solution. 

\textit{Graph Heuristics:} \textsc{Lp-Quts} leverages the complementary slackness condition to construct a reduced graph $G_{RLP}$ from the original graph $G$ by retaining only those edges $(i, j) \in E$ that are tight in the current RLP solution ($\pi_{ij}^{\ast} > 0$). The reduced graph encodes the combinatorial structure of the current RLP. Geometrically, the RLP defines a convex polytope in the space spanned by $\{x_i\}$, which contains all fractional and integer feasible solutions allowed by the constraints. Tight edges define facets of this polytope, and thus determine the local geometry of the current optimal region. Because the construction of $G_{RLP}$ only removes edges from $G$, it is generally sparser (see Fig.~\ref{fig:perfideal} a)) and easier to embed on an atomic register of a NAQC (see steps 3 and 4 in Fig.~\ref{fig:algo}).

\textit{Quantum Sampler:} \textsc{LP-Quts} uses a quantum independent-set sampler adapted to a NAQC to generate $M$ samples of the reduced graph $G_{RLP}$ (step 4 of Fig.~\ref{fig:algo}). These samples are subsequently lifted to independent sets of the original graph $G$ via the maximalize greedy post-processing procedure described in Appendix B of \cite{perron2025leveraginganalogneutralatom} (see step 5 of Fig.~\ref{fig:algo}). This is a single bitflip post-processing technique designed to correct independent set violations by removing the low-weight conflicting vertices and greedily add the highest-weight free vertices. It runs in $O(\text{poly}(|V|))$.

In NAQCs, neutral alkaline atoms are trapped in a register consisting of arbitrary positions $\Vec{r}_i$ in a two-dimensional plane. Once they are cooled to their ground states and initialized in the product state $\ket{0}^{\otimes N}$, the system undergoes analog evolution under a time-dependent Rydberg Hamiltonian to a desired final quantum state:
\begin{equation}
    |\phi(T)\rangle = \mathcal{T} \left[ \exp \left( -i\int_{t=0}^{T} H_{\text{Ryd}}(t) dt\right) \right] \ket{0}^{\otimes N} \label{eq:unitarysimple}
\end{equation}
where where $\mathcal{T}$ is the time-ordering operator, and $H_{\text{Ryd}}(t)$ is comprised of time-dependent global laser fields that coherently drive transitions between the ground state $\ket{g} = \ket{0}$ and a chosen Rydberg excited state $\ket{r} = \ket{1}$ ~\footnote{We choose $|g\rangle = |5S_{1/2}, F=2, m_F = 2\rangle$ and $|r\rangle=|60S_{1/2}, m_J = 1/2\rangle$.} as well as the static van der Walls (vdW) repulsive interactions between the atoms. It can be expressed as $(\hbar = 1)$ \cite{leclerc2024quantum}:
\begin{equation}
    H_{\text{Ryd}}(t) = \frac{\Omega(t)}{2}\sum^N_{i = 1}  \sigma^x_i - \sum^N_{i = 1}\delta_i(t) \hat n_i + \sum_{i < j} \frac{C_6}{r_{ij}^6} \hat n_i \hat n_j
\end{equation}
where $\sigma^x_i = \ket{r}_i \bra{g}_i + \ket{g}_i \bra{r}_i$, $\hat n_i = \ket{r}_i \bra{r}_i$, $\Omega(t)$ is the global Rabi oscillation frequency, $\delta_i(t)$ is the local laser detuning that can be engineered from a global detuning field and a detuning map modulator (DMM). The last term describes the interactions between the Rydberg excited states~\footnote{In our setup, we have $C_6/\hbar \simeq 2 \pi  \times 137 \text{GHz} \cdot \mu\text{m}^6$}, and $r_{ij}$ is the distance between atoms $i$ and $j$. The Hamiltonian at final time $t = T$ encodes the MWIS of $G_{RLP}$ as its ground state by encoding the ratio of the weights of the nodes in the final-time detuning ratios: $\delta_i(T)/\delta_j(T) = w_i/w_j \; \forall i,j \in V$. While other quantum hardware could in principle implement  similar protocols as that of Eq.~\eqref{eq:unitarysimple}, NAQCs are particularly well-suited because the Rydberg blockade naturally enforces the independence constraints \eqref{eq-psp2} with very little overhead~\cite{gibson2025quantum}. To optimize the mapping of the reduced graph $G_{RLP}$ onto the atoms, we determine the atom positions using a Fruchterman-Reingold force-directed algorithm from Ref.~\cite{coelho_efficient_2022}, where edges are treated as springs. We scale the spring constant of each edge by its corresponding dual value $\pi_{ij}$ to ensure that tighter edges in the RLP are represented by closer atomic proximity, and thus deeper within the Rydberg blockade regime. The analog pulse driving $(\Omega(t), \delta(t))$ is implemented using the QSOL strategy of \cite{perron2025leveraginganalogneutralatom} and we choose total time $T = 4 \; \mu\text{s}$ for the entire protocol \footnote{The optimal Rydberg blockade radius $R_b$ for a given register is determined by minimizing the loss function described by Eq.~(14) of \cite{perron2025leveraginganalogneutralatom} with $\lambda = 2$ quantifying the mismatch between the edge structure of the embedding graph of the atomic layout defined by the blockade constraint and $G_{RLP}$.}.

\begin{figure*}[t!]
  \centering
  \includegraphics[width = 0.85\textwidth]{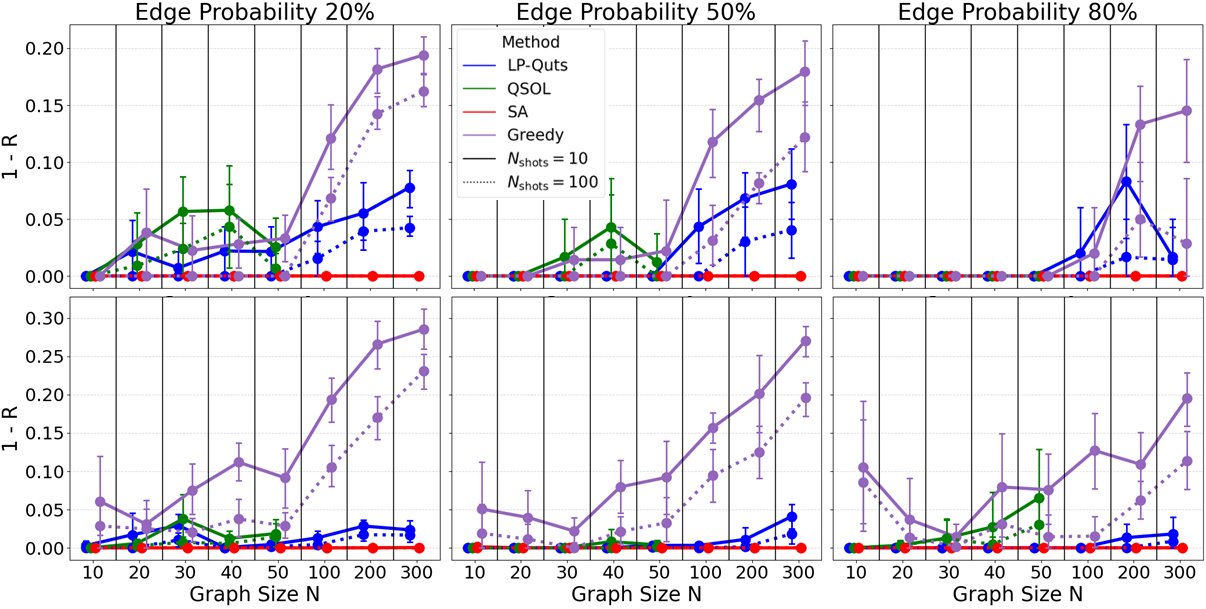}
  \caption{
        Normalized optimality gap $1 - R$, where $R = C_{\mathrm{best}}/C_{\mathrm{opt}}$ is the ratio between the cost of the best sampled solution and the optimal cost $C_{\mathrm{opt}}$, for \textsc{Lp-Quts} and three comparative samplers for the MIS (top) and MWIS (bottom) problems on 10 synthetic Erd\H{o}s--Rényi instances. The optimal cost $C_{\mathrm{opt}}$ is obtained by solving the integer linear program of Eq.~\eqref{eq-psp} using the Gurobi solver~\cite{gurobi}. Each column corresponds to a different edge probability $p$, and each subplot  shows results for varying graph sizes $N$ (x-axis). All methods are compared using the same total sampling budget, matched to that of \textsc{Lp-Quts}, given by $N_{\text{it}} \cdot N_{\text{shots}}$, where $N_{\text{it}}$ is the number of iterations and $N_{\text{shots}}$ is the number of shots per iteration (specified in the legend). \textsc{Lp-Quts} uses at most $\min(N, 40)$ qubits and performs as well as or better than QSOL, which has the same pulse design as \textsc{Lp-Quts} with an improved embedding heuristics applied to the full graph (thus using $N$ qubits). \textsc{Lp-Quts} tackles larger instances beyond the limits of open-source classical emulators ($ \sim 50$ qubits) and current neutral-atom quantum computers (100--200 qubits \cite{wurtz2023aquilaqueras256qubitneutralatom}) achieving significantly superior performance than Greedy for both MIS and MWIS problems and near-optimality for the largest MWIS instances tested, though SA outperforms it at this scale. 
        }
  \label{fig:results}
\end{figure*}

\textit{Cutting-Plane:} \textsc{Lp-Quts} then uses the samples obtained from step 5) of Fig.~\ref{fig:algo} to inform the choice of cuts to the RLP.
As mentioned above, the solution space of the RLP is typically much larger than (and contains) all integer-feasible solutions of Eq.~\eqref{eq-psp}. Cutting-plane algorithms thus progressively shrink this space by introducing new constraints to the RLP that remove fractional solutions whilst not excluding any integer ones. It is  generally a very efficient approach to solving combinatorial optimization problems, even though it still has worst-case exponential complexity with the size of the problem \cite{mahjoub2004_chap_polyedres, papadimitriou1998combinatorial}. In \textsc{Lp-Quts}, we consider odd-cycle constraints, though other types of constraints can also be considered. Let $C$ be an odd cycle containing $\abs{C} = 2k + 1$ vertices. Any independent set can contain at most $k$ vertices of $C$, and therefore respects the inequality:
\begin{equation}
    \sum_{i \in C} x_i \leq \frac{\abs{C} - 1}{2} \label{eq:odd-cycle-ineq}
\end{equation}
where $C$ is a cycle of odd length and $x_i$ is the value assigned to the vertex $i$ in the integer solution. The above inequality can be violated by fractional solutions of the RLP. For example, a standard fractional solution is where $x_i = 0.5$ for all vertices in a odd cycle of length $\abs{C} = 2k +1$. Then, the above inequality gives: $k + 0.5 \leq k$, which is a violation. The separation problem then consists in identifying the odd cycle $C^*$ with the largest violation to the constraint \eqref{eq:odd-cycle-ineq} given a RLP solution $\mathbf{x}^{RLP}$:

\begin{equation}
    \max_{C*} \left ( \varepsilon_{RLP}(C^*) =  \sum_{i \in C^*} x^{RLP}_i  -(\abs{C^{*}} - 1)/2 \right ) \label{eq:sepdjik}
\end{equation}

Although the number of odd cycles scale exponentially with the size of the graph, Eq.~\ref{eq:sepdjik} can solved in time polynomial with $|V|$ and $|E|$ using a modified Djikstra algorithm \cite{mahjoub2004_chap_polyedres, grotschel1988geometric} (see Appendix \ref{app:oddcyclesep} for details). This algorithm generates $\abs{V}$ (potentially identical) odd cycles. In many practical instances, the separation problem often has a large number of odd cycles with identical maximal violation. These cycles are equally valid from the standpoint of classical separation, but may vary considerably in how useful they are for tightening the relaxation. Intuitively, cuts lying closer to the integer hull, and more importantly, in the vicinity of higher-weight integer solutions, tend to provide a stronger progress as they remove a more substantial portion of the fractional region in directions that are actually relevant for finding the optimal solution. We provide an illustrative example in step 6) of Fig.~\ref{fig:algo}. Motivated by this, we define a modified separation problem, which conceptually induces a Pareto-style trade-off between two objectives: maximizing classical violation and maximizing proximity of the cuts to the integer solutions sampled. This results in the modified separation problem:

\begin{subequations} \label{eq-sep}
\begin{align}
    \argmax_{C^*} & \left[ \varepsilon_\alpha(C^*) =  \varepsilon_{RLP}(C^*) + \alpha \varepsilon_{\mathcal{S}}(C^*) \right] \label{eq-sep1} \\
 \varepsilon_{\mathcal{S}}(C^*) &=  \sum_{i \in C^*}  \langle n_i \rangle_{\mathcal{S}^{t}}  -  \frac{(|C^*| -1)}{2} \label{eq-sep1-2} \\
    & \text{s.t.} \sum_{i \in C^*} x^{RLP}_i - \frac{\abs{C^*} - 1}{2} > 0
    \label{eq-sep2}
\end{align}
\end{subequations}
where $C^*$ is an odd-cycle, $\langle n_i \rangle_{\mathcal{S}^{t}}$ denotes the average occupation of vertex $i$ over the set $\mathcal{S}^{t}$ of independent sets sampled at iteration $t$, and $\alpha \geq 0$ is a weighting parameter that controls the relative influence of the sampler-informed term on cut selection. Particularly, it is gradually tuned from $\alpha_{\text{max}} > 0$ down to $0$ to ensure that all odd-cycle constraints violating the current RLP solution, i.e. respecting constraint \eqref{eq-sep2}, are found and added. The algorithm used in \textsc{Lp-Quts} to tune $\alpha$ is described in Appendix \ref{app:alpha_tuning}. This modified separation problem can still be solved efficiently using the same Dijkstra's algorithm as in the classical separation problem, though the solution now carries information from the sampled solutions and lifts possible degeneracies in found cuts. In Fig.~\ref{fig:perfideal} b), we show that this modified separation problem converges to the optimal solution faster for series-parallel graphs than the traditional one (given by $\alpha=0$). Once the most violating cycle $C^*$ is found, its corresponding odd-cycle inequality \eqref{eq:odd-cycle-ineq} is added as a constraint to the RLP. Moreover, a new dual variable $\mu_e \geq 0$ for each edge $e \in C^*$ is introduced in the dual problem \eqref{eq:dual}, and contributes a term $\mu_e (\abs{C^*} - 1)/2$ to the dual objective, and has an associated dual constraint $\sum_{e \in C^*} \mu_e \geq 0$. Since Dijkstra's algorithm identifies $\abs{V}$ odd cycles in one implementation, more than one violating odd cycles can be added to the RLP at each iteration. In \textsc{Lp-Quts}, all violating odd cycles found are added.

For a certain class of graphs known as t-perfect graphs \cite{chudnovsky2024colouringtperfectgraphs, bruhn2012clawfree, EisenbrandEtAl2003Tperfect}, the convex hull of all integer solutions can be described exactly by the independence constraints of Eq.~\eqref{eq-psp3}, non-negativity constraints  ($0 \leq x_i \leq 1 \; \forall i \in V$), and the odd-cycle constraints  ~\eqref{eq:odd-cycle-ineq} \cite{mahjoub2004_chap_polyedres, grotschel1988geometric}. These constraints are thus \textit{facet-defining} for these graphs, and the above cutting-plane approach is guaranteed to converge to the optimal solution in polynomial time with the size of the graph. For arbitrary graphs, however, these constraints alone are generally insufficient to guarantee convergence, and the cutting-plane algorithm must necessarily rely on additional inequalities \cite{disser_clique, grotschel1988geometric} whose separation problem is known to be NP-hard. We demonstrate in Fig.~\ref{fig:perfideal} c) that the number of iterations $N_{\rm iters}$ to converge for a certain class of t-perfect graph called series-parallel graphs \cite{EisenbrandEtAl2003Tperfect} increases polynomially with the size of the graph.

Finally, \textsc{Lp-Quts} operates via a reinforcing feedback loop: first the RLP is solved to provide the upper bound and the structure of the reduced graph $G_{RLP}$ from which independent sets are sampled from. The best independent set thus sampled provides a lower bound, while the set of samples $\mathcal{S}$ improves the selection of the odd-cycle constraints to tighten the RLP. This improves the upper bound  and alters the structure of $G_{RLP}$ to keep all tight inequalities that describe the current RLP. This produces higher-quality samples; ipso facto, the lower bound is increased. Iterating this loop thus narrows the gap (see step 7 of Fig.\ref{fig:algo}) and guides the algorithm towards the optimal solution.

\section{Results and Discussion}
\label{sec:results-discussion}

We first compare \textsc{Lp-Quts} to heuristic solvers operating directly on the full graph, without benefiting from \textsc{Lp-Quts}' decompositions. We implement a classical greedy solver, a simulated annealing (SA) solver and an analog quantum solver (QSOL). All solvers are followed by the maximalize greedy post-processing routine of \cite{perron2025leveraginganalogneutralatom}, which is also used by \textsc{Lp-Quts} in step 5). The greedy solver constructs an independent set by inserting vertices according to probabilities proportional to the cumulative distribution of their (normalized) weights. Different solutions are obtained by randomizing the order of the considered vertices. The SA solver is implemented using the \textsc{Neal} Python library \cite{dwave_ocean_samplers}, where each instance is encoded as a quadratic unconstrained binary optimization (QUBO) $Q_G \;=\; -\sum_{i\in V} \bar{w}_i\, n_i + 2\sum_{(i,j)\in E} n_i n_j $ with normalized weights $\bar{w}_i$ and a penalty factor $2$ chosen to dominate any pairwise weight sum $\bar{w}_i + \bar{w}_j$. The annealing schedule uses inverse temperatures $\beta_i = 0.01$ and $\beta_f = 100$, with a fixed number of steps. The quantum solver (QSOL) uses the same analog pulse schedule as \textsc{Lp-Quts}, but its embedding is generated using the \textsc{SA-embedder} heuristic of Ref.~\cite{perron2025leveraginganalogneutralatom} with the default hyperparameters reported therein. This heuristic typically yields higher-quality embeddings than the force-directed method used in \textsc{Lp-Quts}, albeit at a larger classical computational cost. 

\begin{figure}[t!]
  \centering
  \includegraphics[width = 0.485\textwidth]{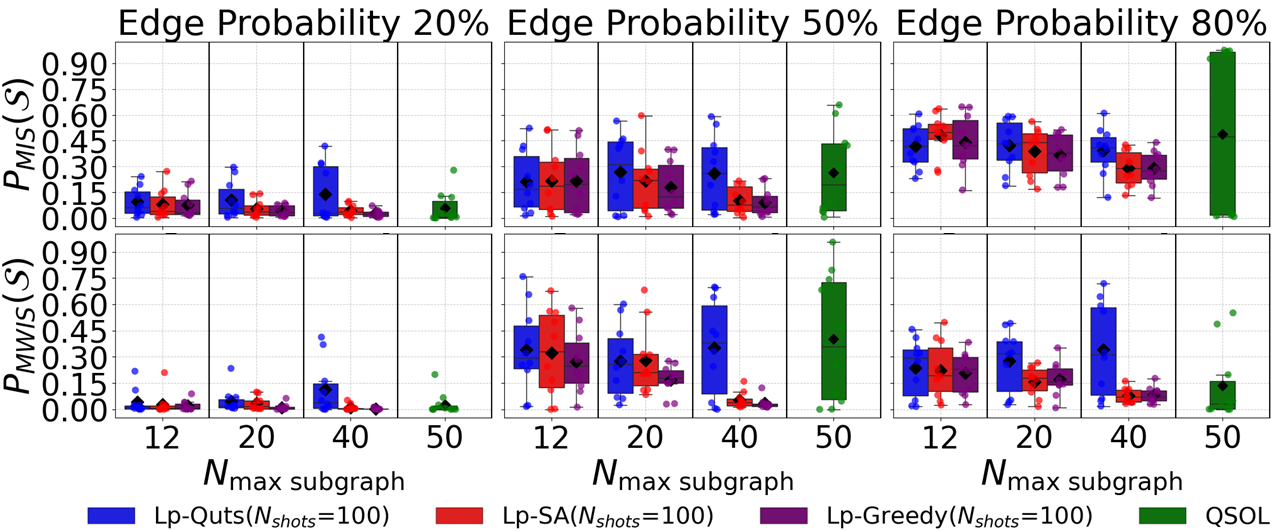}
  \caption{
  Probability $P_{\text{MIS}}$ (top) and $P_{\text{MWIS}}$ of obtaining the optimal solution among the sampled solutions  (bottom) versus the maximum subgraph size for 10 random MIS and MWIS instances of size $N=50$ with varying edge density (columns). We run \textsc{Lp-Quts} (blue) and compare it with \textsc{Lp-Greedy} (purple) and \textsc{Lp-SA} (red) to assess whether comparable solution quality can be achieved purely classically. All methods use $N_{\text{shots}} = 100$ shots per iteration. As an additional benchmark, we include QSOL (green) on the full graph ($N_{\text{max-qubits}} = 50$), adjusted to use the same total number of samples as  \protect\textsc{Lp-Quts}. We show the individual instances as scatter points and the mean with a black diamond. Consistently,  \protect\textsc{Lp-Quts} outperforms its classical counterparts.
  }
  \label{fig:internal}
\end{figure}

All quantum protocols in QSOL and \textsc{Lp-Quts} are implemented using the \textsc{Pulser} Python package. Emulations are performed using the \textsc{QuTip} backend \cite{johansson2012qutip} for $N \leq 12$ qubits, the \textsc{emu-sv} backend \cite{bidzhiev2025efficientemulation} for $12 < N \leq 27$ qubits, and the \textsc{emu-mps} backend \cite{bidzhiev2025efficientemulation} for $N > 27$ qubits, with the maximum bond dimension of the matrix product state set to $\chi = 512$ in the latter case. Aside from this modification, all emulator parameters are kept at their default values and all simulations are performed in the absence of noise. Unless otherwise specified, we limit the number of qubits in \textsc{Lp-Quts} to $N_{\text{max-qubits}} = \min(N, 40)$, where $N$ is the graph size, to control memory requirements needed to ensure that the emulations maintain the desired precision. Clusters of $G_{RLP}$ with size greater than $N_{\text{max-qubits}}$ are partitioned by iteratively removing the edge $e = (i,j)$ from those clusters with the smallest dual value $\pi_{ij}$ until the largest size of the resulting partitions are less than this limit. Finally, we set the number of samples taken at each iteration to $N_{\text{shots}}$, which we implicitly write in the form \textsc{Lp-Quts}($N_{\text{shots}}$) in figure legends. We set the maximum number of iterations of \textsc{Lp-Quts} to 20 and stop the algorithm if no improvements were identified after 4 consecutive iterations.

\textit{Results:} We evaluate the performance of \textsc{Lp-Quts} on connected Erd\H{o}s--Rényi graphs with $N$ vertices and edge density $p$, considering sizes up to $N = 300$. For each pair $(N, p)$, we generate 10 random instances of MIS and MWIS problems. In the MWIS case, vertex weights are drawn independently from an uniform distribution $\mathcal{U}(0,1)$. For each instances, the optimal solution was found by solving the integer linear program of Eq.~\eqref{eq-psp} using the Gurobi solver~\cite{gurobi} (version 12.0.0) on 5 CPU cores (Intel 6972P @ 2.4 GHz) with a maximum memory of 50GB and a maximum runtime of 5 hours.

Fig.~\ref{fig:results} shows the main results of \textsc{Lp-Quts}. It reports the gap to the optimal solution for Erd\H{o}s--Rényi instances of MIS and MWIS problems, comparing \textsc{Lp-Quts} against our benchmarks under an equal sampling budget per instance. \textsc{Lp-Quts} performs better compared to the QSOL solver, even in regimes where it uses fewer total qubit resources (e.g. for $N = 50$). Importantly, \textsc{Lp-Quts} continues to deliver solutions within less than 10\% and 5\% of optimality at worst for MIS and MWIS instances respectively for larger graphs that lie beyond the reach of current open-source NAQCs emulators ($N_{\text{qubits}} \sim 50$) \cite{bidzhiev2025efficientemulation} and current NAQCs devices ($N_{\text{qubits}} \sim 100-300$) \cite{wurtz2023aquilaqueras256qubitneutralatom, PasqalOrion}. These results highlight the ability of \textsc{Lp-Quts} to extract strong performance while relying on limited quantum resources. We also observe that \textsc{Lp-Quts} tends to perform better on MWIS than on MIS problem, a trend that does not hold for QSOL (for instance at $p = 80$ \%). This behavior can be explained by the structure of the RLP solution; in MWIS, vertex weights bias fractional values so that high-weight vertices often satisfy $x^{RLP}_i > 0.5$ even at the first iteration, whereas in MIS, the RLP generally starts with uniform fractional values $x^{RLP}_i = 0.5$. As a result, the reduced graph $G_{RLP}$ generally tends to more closely approximates the integer hull in MWIS instances; hence providing better samples. 

Compared with classical solvers, \textsc{Lp-Quts} significantly outperforms the greedy solver across nearly all tested graph sizes, with the performance gap widening as graph size increases. However, it performs worse than classical simulated annealing (SA), which consistently samples the optimal solution of all instances under the same sample budget. Observing a degradation in SA's performance would require testing substantially larger graphs, but this will also demand larger quantum resources than the $N_{\text{max-qubits}} \approx 50$ limit imposed by open-source emulators as will be discussed shortly. Pushing our benchmark further while quantifying where SA may fail remains a challenge and is part of ongoing work. Note that, for t-perfect graphs, neither greedy, SA or QSOL have guarantees. \textsc{Lp-Quts}, on the other hand, inherits guarantees from operations research, which results in a convergence to optimality in polynomial time for this class.

The gap to optimality also depends strongly on the total sampling budget, as seen in the dashed curves of Fig.~\ref{fig:results}. The performance of \textsc{Lp-Quts} is immediately improved by increasing the sampling budget. This is inherent to the NP-hardness of MIS and MWIS: for any polynomial-time sampler, recovering the optimal solution with high probability generally requires exponentially many samples. A practical way to improve \textsc{Lp-Quts}  is to regularly increase the number of samples taken at each subsequent iteration. As additional constraints are found and added, $G_{RLP}$ tends to be more representative, and sampled solutions are of higher quality. It is therefore better to dedicate costly sampling budget to later iterations.

We next study the impact of replacing the quantum solver of step 4 of \textsc{Lp-Quts} with greedy (\textsc{Lp-Greedy}) and simulated annealing (\textsc{Lp-SA}) methods as the available resources ($N_{\text{max subgraph}}$) is increased. As seen in Fig.~\ref{fig:internal},  \textsc{Lp-Quts} consistently outperforms its classical counterparts, with the performance gap widening as the maximum subgraph size $N_{\text{max subgraph}}$ increases. While \textsc{Lp-Quts} improves with increasing resources, the performance of \textsc{Lp-greedy} and \textsc{Lp-SA} degrades. We attribute this to the quantum sampler’s ability to generate samples that are both high quality and diverse, in contrast to greedy sampling (high diversity, low quality) and SA (high quality, low diversity), as seen also in Ref.~\cite{perron2025leveraginganalogneutralatom}. This combination enables \textsc{Lp-Quts} to explore a broader low-energy landscape and prevents the greedy post-processing from augmenting the samples into the same independent sets (i.e. local minimas) of the original graph.

When compared with QSOL, \textsc{Lp-Quts} samples the optimal MIS/MWIS solution(s) more frequently for sparse graphs $(p = 20 \%)$ and achieves comparable performance in denser regimes. Interestingly, 
in denser regimes, \textsc{Lp-Quts} more consistently samples the optimal solutions at least once across most instances, while QSOL shows a more uneven performance, where it samples the optimal solutions frequently for some instances, but in other instances missing it entirely. This points to the fact that  \textsc{Lp-Quts} benefits from sampling over multiple $G_{RLP}$ hence multiple embeddings, whereas QSOL samples on the same embedding, irrespective of its quality.

The amount of resources allocated $N_{\text{max subgraph}}$ is critical. It should notably scale with the size of the graph $\mathcal{O}(N)$. In late iterations, $G_{RLP}$ decomposes into one or few large clusters of size $\mathcal{O}(N)$ along with many smaller clusters of size $\mathcal{O}(1)$. Performance differences between samplers are therefore mostly driven by how effectively they sample independent sets within these large clusters. At $N_{\text{max subgraph}} = 12, 20$, the dominant clusters become too fragmented and the potential advantage that  \textsc{Lp-Quts}'s quantum subroutine may provide is largely lost. Additional supporting evidence is provided in Appendix \ref{app:SM1}, where for $p = 20 \%$ a modest advantage in the quality of the samples generated by \textsc{Lp-Quts} is observed over those generated by \textsc{Lp-SA} and \textsc{Lp-Greedy} at $N = 100$ at the largest $N_{\text{max subgraph}} = 40$, but this advantage disappears for $N = 200$ and $N = 300$ under the same resource constraint. Thus, exploring larger instances, which in turns  requires the ability to deploy \textsc{Lp-Quts} at scale on NAQC hardware, is critical for further characterization of its performance.

\begin{figure}[t!]
  \centering
  \includegraphics[width = 0.485\textwidth]{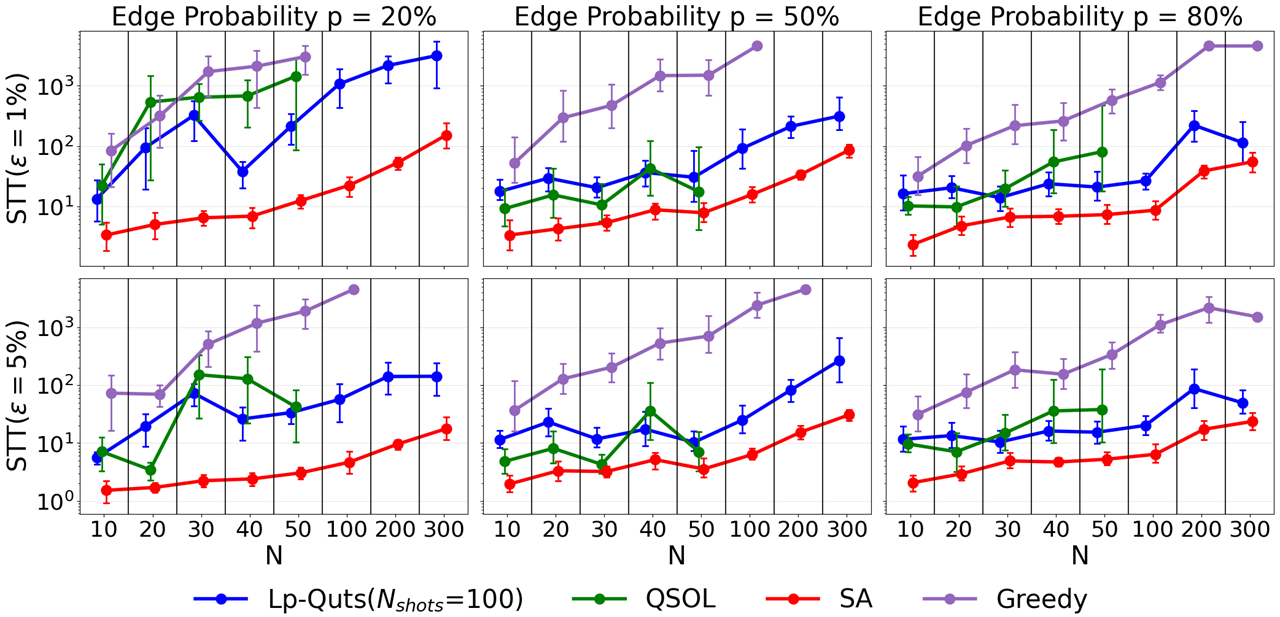}
  \caption{Sample-to-target (STT($\epsilon$)) energy as a function of graph size $N$ and edge density $p$ (columns) for MWIS instances. We compare \textsc{Lp-Quts} and the three classical solvers over the same total sample budget for each instance. The first and second rows correspond to $\epsilon = 1\%$ and $\epsilon = 5\%$, respectively. Instances in which no sample reached the target energy are omitted. To ensure fair comparisons, the averages for QSOL, \textsc{Lp-Quts}, and SA are computed over instances where all methods return well-defined STT values. 
  }
  \label{fig:stt}
\end{figure}

We benchmark \textsc{Lp-Quts} using the Sample-to-Target energy (STT($\epsilon$)) metric, which measures the number of samples required to obtain, with 99\% confidence, a solution within $\epsilon\%$ of the optimal energy, defined as:
\begin{equation}
STT(\epsilon) = \frac{\log(1 - 0.99)}{\log\left(1 - p_{E \leq E_0 ( 1 - \epsilon)}\right)}
\end{equation}
These results are presented in Fig.~\ref{fig:stt} for MWIS instances. We omit showing a similar plot for MIS instances due to the highly discrete cost structure (if $\epsilon < 1/|\text{MIS}|$, STT $\rightarrow \infty$ unless the optimal solution was found). We first observe that \textsc{Lp-Quts} achieves STT values comparable to those of QSOL across the system sizes considered. This behavior reflects the fact that, in this regime, QSOL operates under relatively mild embedding errors, which can still be effectively mitigated through simple single-bit-flip greedy post-processing. As a result, both approaches exhibit similar sampling efficiency at these scales. Second, as expected from the NP-hardness of the MIS and MWIS problems, STT increases exponentially with system size for all methods. Among the four methods tested, SA consistently achieves the lowest STT, generally outperforming the second best \textsc{Lp-Quts} by a factor of 10. Nevertheless, our results suggest that \textsc{Lp-Quts} may achieve more favorable asymptotic scaling. A detailed investigation of this behavior is essential and is part of ongoing work.

\section{Conclusion}
\label{sec:conclusion}

In this work, we introduced \textsc{Lp-Quts}, an iterative framework that combines a cutting-plane algorithm with a quantum sampler implemented on neutral-atom quantum computers to tackle the maximum-weight independent set (MWIS) problem. We benchmarked \textsc{Lp-Quts} against classical baselines, including simulated annealing and a weighted greedy heuristic, as well as against an analog quantum solver applied directly to the full problem graph using the same analog pulse schedule but a more powerful embedding heuristic \cite{perron2025leveraginganalogneutralatom}. \textsc{Lp-Quts} achieves superior to equal performance on the same sampling budget to the latter quantum sampler even on problems where it has access to less qubit resources. More importantly, \textsc{Lp-Quts} tackles problems of much larger size than currently accessible by either NAQC emulators \cite{bidzhiev2025efficientemulation} and current hardware \cite{wurtz2023aquilaqueras256qubitneutralatom, PasqalOrion}, all the while giving solutions within 5-10\% of optimality for those specific instances. While it does not outperform simulated annealing, which is capable, under the same sampling budget as \textsc{Lp-Quts}, to sample the optimal solution of all instances tested, our empirical studies of the sample-to-target energy suggests that \textsc{Lp-Quts} may exhibit more favourable scaling at the larger problem sizes tested ($N = 200-300$). An important future direction will be to more sharply characterize this trend, especially as both problem sizes and available quantum resources grow.

More broadly, \textsc{Lp-Quts} illustrates how quantum heuristics can be systematically integrated with classical operational research (OR) tools, such as linear relaxation and cutting planes, rather than being applied in isolation. In particular, \textsc{Lp-Quts} inherits performance guarantees on t-perfect graphs from classical OR methods, which we explicitly show for an important subclass of t-perfect graphs known as series--parallel graphs \cite{EisenbrandEtAl2003Tperfect} (see Fig.~\ref{fig:perfideal} c). On these "easy" instances, we even show that \textsc{LP-Quts} benefits from the sample-informed separation method, resulting in a modest acceleration (see Fig.~\ref{fig:perfideal} b). As such, our hybrid algorithm directly addresses a key limitation of many quantum heuristics for MWIS, which do not have any performance guarantees on classically-easy instances. We hope this result further pushes the community to embrace and explore the marriage of OR tools with quantum techniques.

Moreover, our framework allows us to naturally formulate a graph heuristic where $G_{\rm RLP}$ is obtained from the solution of the relaxed linear program. This graph exhibits a dramatically sparser structure than the original problem graph. In particular, for the largest instances considered ($N = 300$), we note a 2 orders of magnitude reduction on average in the number of edges (see Fig.~\ref{fig:perfideal} a). Importantly, we show that sampling independent sets from this reduced graph with our quantum sampler, then uplifting those samples to independent sets of the problem graph yields independent sets of similar to better quality than sampling directly from the problem graph and improving the samples with the same post-processing. Incidentally, the sparseness of the reduced graph greatly improves the quality of the resulting embedding into the atomic register, which mitigates a major challenge in neutral-atom quantum computing that the hardware connectivity is constrained by a fixed unit-disk (UD) geometry, which is generally incompatible with the geometry of MWIS problems relevant in industry.

\acknowledgments

We thank Wesley Coelho, Yassine Naghmouchi and
Hossein Sadeghi for valuable discussions and support. This research was financially supported by the Natural Sciences and Engineering Research Council of Canada (NSERC) and Pasqal Canada through an Alliance grant (ALLRP 590810 - 23). This work made use of compute resources by Calcul Québec and the Digital Research Alliance of Canada.

\appendix
\section{Separation Algorithm for Odd-Cycle Constraints}
\label{app:oddcyclesep}
We present the algorithm (Algorithm \ref{alg:oddcyclesep}) used to solve the separation problem for the odd-cycle constraints ~\eqref{eq-sep} for a fixed parameter $\alpha \ge 0$ on the problem graph $G$. The algorithm reduces the separation problem to shortest-path computations of a bipartite graph $B_G$ obtained from $G$. For each vertex of the original graph, a shortest-path search is performed between its two copies in the bipartite graph. The resulting path of $B_G$ yields an odd cycle of $G$, thus the full implementation of the algorithm yields $\abs{V}$ potentially degenerate odd cycles. The shortest-path search can be efficiently carried out using Dijkstra’s algorithm using \textsc{NetworkX} Python package \cite{networkx_dijkstra}. This algorithm runs in $\mathcal{O}(|V|\log |V| + |E|)$ for a single computation. Since this procedure is repeated for all vertices, the overall complexity of
the algorithm is $\mathcal{O}(|V|^2 \log |V| + |V| \cdot |E|)$.

\begin{algorithm}[htbp]
\caption{Separation of Odd-Cycle Constraints using \texorpdfstring{\textsc{Dijkstra}}{Dijkstra}}
\label{alg:oddcyclesep}
\KwIn{Graph $G=(V,E)$, RLP solution $x^{RLP}$, sampled independent-set statistics $\langle n_i \rangle_{\mathcal{S}^t}$, parameter $\alpha \ge 0$}
\KwOut{Set $\mathcal{C}$ of violated odd-cycle constraints}

Initialize $\mathcal{C} \gets \emptyset$\;

\textbf{Edge weight definition:}\;
\ForEach{$(i,j) \in E$}{
    $z(i,j) \gets (1 - x^{RLP}_i - x^{RLP}_j) + \alpha \, (1 - \langle n_i + n_j \rangle_{\mathcal{S}^t})$\;
}

\textbf{Bipartite graph construction $B_G = (V' \cup V'', \tilde{E})$} \;
\ForEach{$v \in V$}{
    Create vertices $v' \in V'$ and $v'' \in V''$\;
}
\ForEach{$(u,v) \in E$}{
    Add edges $(u',v'')$ and $(u'',v')$ in $\tilde{E}$ with weight $z(u,v)$\;
}

\textbf{Shortest-path search:}\;
\ForEach{$v \in V$}{
    Run \textsc{Dijkstra}$(B_G, v', v'')$ to find the shortest path\;
    Extract minimum-weight path $P_v$ from $v'$ to $v''$ with total weight $W(P_v)$\;
    Project $P_v$ onto $V$ to obtain odd cycle $C_v$ with cost $\varepsilon_\alpha(C_v) = W(P_v)$\;
    \tcp{Projection: if $u'$ or $u''$ in $P_v$, add $u$ to $C_v$}
    \If{$\sum_{i\in C_v} x^{RLP}_i - \frac{|C_v|-1}{2} > 0$}{
        $\mathcal{C} \gets \mathcal{C} \cup \{C_v\}$\;
    }
}

\Return{$\mathcal{C}$}\;
\end{algorithm}

\section{Alpha-Tuning Algorithm for Odd-Cycle Separation}
\label{app:alpha_tuning}

We present the $\alpha$-tuning algorithm (Algorithm~\ref{alg:alpha_schedule}) that iteratively decreases $\alpha$ from $1$ to $0$ in $N_{\text{steps}}$, which is required to ensure the convergence of the RLP for t-perfect graphs. We use Algorithm \ref{alg:oddcyclesep} to find all odd cycles violating Eq.~\eqref{eq-sep2}. The procedure has complexity 
$ \mathcal{O}\Big(N_{\text{steps}} \cdot (|V|^2 \log |V| + |V| \cdot |E|)\Big)$, where $V$ and $E$ are the vertices and edges and $N_{\text{steps}}$ is the total number of steps used to decrease $\alpha$ from $1$ to $0$.

\begin{algorithm}[htbp]
\caption{Iterative $\alpha$-Tuning for Separation Problem}
\label{alg:alpha_schedule}
\KwIn{Graph $G$, RLP solution $x^{RLP}$, sampled independent sets $\mathcal{S}^t$, number of $\alpha$ steps $N_{\text{steps}}$}
\KwOut{Candidate odd cycles to add to RLP}

Initialize $\alpha_0 \gets 1$\;
Set step size $\Delta \alpha \gets \alpha_0 / N_{\text{steps}}$\;
Set $\alpha \gets \alpha_0$\;
Initialize candidate set $\mathcal{C} \gets \emptyset$\;

\While{$\alpha \ge 0$}{
    Run Algorithm~\ref{alg:oddcyclesep} with current $\alpha$ to identify odd cycles\;
    \If{no new cycles were found}{
        Update $\alpha \gets \alpha - \Delta \alpha$\;
    }
    \Else{
        Add the identified cycles to $\mathcal{C}$\;
        \textbf{break}\;
    }
}

\Return{$\mathcal{C}$}\;
\end{algorithm}

\section{Effect of Maximum Subgraph Size on Sampler Performance}

\begin{figure}
  \centering
  \includegraphics[width = 0.485\textwidth]{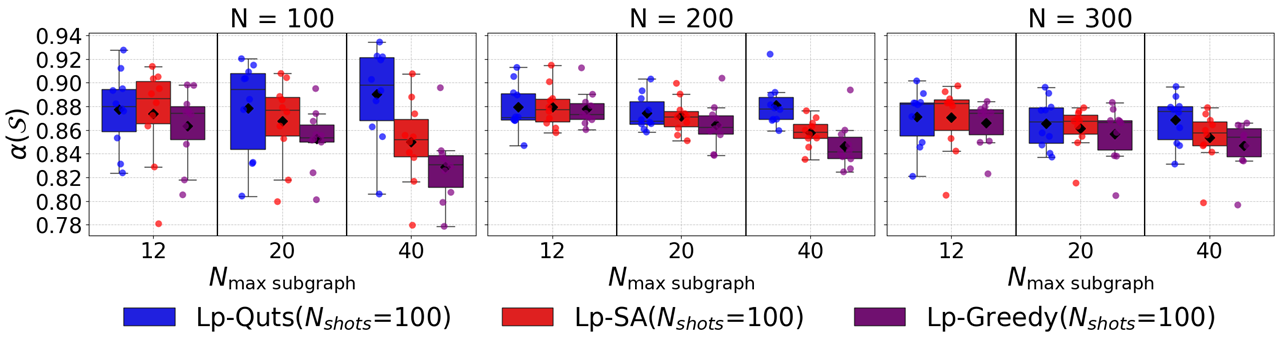}
  \caption{
  Approximation ratio $\alpha(\mathcal{S})$ versus the maximum subgraph size for 10 random MIS and MWIS instances of varying sizes (columns) and edge density $p = 20\%$. We run \textsc{Lp-Quts} (blue), \textsc{Lp-Greedy} (purple) and \textsc{Lp-SA} (red) to assess whether comparable solution quality can be achieved purely classically. All methods use $N_{\text{shots}} = 100$ per iteration. Individual instances are shown as scatter points and the mean with a black diamond.
  }
  \label{fig:internal2}
\end{figure}

In Fig.~\ref{fig:internal2}, we provide complementary results to those presented in Fig.~\ref{fig:internal}. Here, we show the effect of changing the size of the largest allowed subgraph ($N_{\rm max subgraph}$, i.e. the allowed resources) on the approximation ratio. For an ensemble of independent sets $\mathcal{S}$, the approximation ratio is obtained as

\begin{equation}
    \alpha(\mathcal{S}) = \frac{1}{|\mathcal{S}|} \sum_{s\in \mathcal{S}} \frac{C(s)}{C_{\rm opt}}
\end{equation}

where $C(s)$ is the cost of a given independent set, and $C_{\rm opt}$ is the optimum. This characterization informs on the overall quality of \textit{all} the samples - the higher it is, the better. We see from Fig.~\ref{fig:internal2} that, for graphs with $N = 100$, a modest advantage of \textsc{Lp-Quts} over \textsc{Lp-SA} and \textsc{Lp-Greedy} is observed at the largest tested $N_{\text{max subgraph}} = 40$. This advantage increases as $N_{\text{max subgraph}}$ increases. However, for larger systems ($N = 200$ and $N = 300$), this advantage disappears under the same resource constraints because $N_{\text{max subgraph}} = 40$ is too small to adequately sample the largest clusters of $G_{RLP}$, leading to fragmentation that diminishes the relative performance advantage.

\label{app:SM1}




%
\bibliography{ref-jabref}

\end{document}